Gravitoelectrodynamics in Saturn's F Ring: Encounters with Prometheus and Pandora

Lorin Swint Matthews and Truell W. Hyde
Center for Astrophysics, Space Physics, and Engineering Research
Baylor University
P.O. Box 97310, Waco, Texas, USA 76798-7310

Short title: Gravitoelectrodynamics in Saturn's F Ring

Classifications numbers:
52.27.Lw   Dusty or complex plasmas; plasma crystals
25.52.2b   Complex (dusty) plasma
96.30.Mh   Saturn  (Planets, their satellites, and rings; asteroids)
96.30.Wr   Planetary rings

Abstract
The dynamics of Saturn's F Ring have been a matter of curiosity ever since Voyagers 1 and 2 sent back pictures of the ring's unusual features. Some of these images showed three distinct ringlets with the outer two displaying a kinked and braided appearance.  Many models have been proposed to explain the braiding seen in these images; most of these invoke perturbations caused by the shepherding moons or km sized moonlets imbedded in the ring and are purely gravitational in nature. These models also assume that the plasma densities and charges on the grains are small enough that electromagnetic forces can be ignored. However, Saturn's magnetic field exerts a significant perturbative force on even weakly charged micron and sub-micron sized grains causing the grains to travel in epicyclic orbits about a guiding center. This study examines the effect of Saturn's magnetic field on the dynamics of micron-sized grains along with gravitational interactions between the F Ring's shepherding moons, Prometheus and Pandora.  Due to differences in the charge-to-mass ratios of the various sized grains, a phase difference between different size populations is observed in the wavy orbits imposed by passage of the shepherding moons.

1. Background

Saturn's F Ring is one of the most dynamic rings observed in the solar system. While Voyager 1 images showed three distinct ringlets with the outer two displaying a kinked and braided appearance [1], Voyager 2 images showed a much more regular structure with four separate, non-intersecting strands [2]. Both Voyager 1 and Voyager 2 detected brighter clumps within the rings with a temporal dependence ranging from days to months [3].

Photometric data collected by Voyager 1 and Voyager 2 suggested that the general structure of the F Ring consisted of a core of centimeter to millimeter sized particles ~1 km in width, with an envelope of micron and submicron sized grains extending inward ~50 km [4]. Data collected from the Hubble Space Telescope in 1995 showed similar features, but resulting fits to the data indicated that the minimum grain size is likely to fall within the range of 0.5 to 1 μm, with a maximum of 28% of the ring material being submicron in size [5].

The F Ring lies within the inner part of Saturn's magnetosphere, which contains plasma that rigidly corotates with the planet [6]. Thus the dust in Saturn's F Ring can become charged. Since the plasma parameters in the vicinity of the F Ring are poorly constrained, the magnitude of the charge on the dust grains is not well known. Estimations of the grain charge for micron and submicron grains within the F Ring have ranged from less than one electron [7] to –38 V [8]. The grain charge is of interest since charged grains' orbits will be perturbed by the planetary magnetic field with the magnitude of this perturbation depending primarily on the grain's charge-to-mass ratio. Recently it has been shown that significant perturbations are possible for even weakly charged grains. This allowed the observed physical characteristics of the F Ring, such as the 50 km width of the F Ring envelope, to be used to determine that the micron and submicron grains within the envelope are likely weakly negatively charged, with a maximum $|q/m|$ = 0.03 C/kg [9].

The purpose of this study is to examine the interaction of such weakly charged individual grains with the F Ring's shepherding moons, Prometheus and Pandora, and compare their orbits to that of uncharged grains. Complications not addressed in this study are the inclination of the F Ring and satellite orbits, collisional damping between ring particles, collective effects of the charged grains and plasma, and the effect of moonlets imbedded within the ring.

2. The box_tree algorithm

The numerical code employed in this study is a modified version of box_tree, a self-consistent N-body code developed to study Saturn's rings, planetesimal dynamics, and fractal aggregation [10,11,12]. The current version has been modified to consider charged grains and the electrostatic and magnetic forces acting on them as well as external gravitational potentials such as those produced by shepherding satellites [13]. The code has already been used to model a variety of astrophysical and laboratory dusty plasma environments, including formation of and wave dispersion in dust crystals [14,15], chondrule formation [16], and coagulation in protoplanetary disks [13].

The box code [17] divides the ring into self-similar patches or boxes orbiting the planet where the box size is much greater than the mean radial excursions of the constituent particles. This allows the boxes to be dynamically independent with more distant regions of the ring represented by copies of the simulated region. In a rotating system, particle motions are referenced to the center of a box that orbits the planet at the local Keplerian velocity. The box code, in providing the external potentials acting on the grains, specifies a coordinate system, the linearized equations of motion, and a prescription for handling boundary conditions. The tree code [18] provides a mechanism for the rapid calculation of the interparticle forces by means of a multipole expansion, reducing the CPU scaling time from $O(N^2)$ to $O(N\log N)$ for sufficiently large N. The interparticle forces can then be included as a perturbation to the equations of motion.

2.1 Linearized Equations of Motion

The acceleration of a charged grain in a planetary magnetic field, **B**, and gravitational field is given by

$$\ddot{\mathbf{R}} = \frac{q}{m}(\mathbf{E} + \dot{\mathbf{R}} \times \mathbf{B}) - \frac{GM_p}{R^3}\mathbf{R} - \nabla\phi \qquad (1)$$

where $M_p$ is the mass of the planet, **R** is the distance from the center of the planet to the grain, and *q* and *m* are the grain's charge in coulombs and mass in kilograms, respectively. The grain-grain gravitational and electrostatic interactions are included in the $-\nabla\phi$ term. The electric field results from the relative motion of the ring particles with respect to the corotating magnetic field and is given by

$$\mathbf{E} = -(\mathbf{\Omega_p} \times \mathbf{R}) \times \mathbf{B}(\mathbf{R}) \qquad (2)$$

where $\mathbf{\Omega_p}$ is the angular velocity of the planet.

The equations of motion can be linearized in the box frame, which is rotating about the planet with constant angular velocity $\mathbf{\Omega_k} = \Omega_k \hat{z}$ with magnitude

$$\Omega_k = \sqrt{\frac{GM}{R_b^3}}. \qquad (3)$$

$R_b$ is the distance from the box to the planet center and M is the mass of the central planet. Using the fact that $R_b \gg r$, the position of the particle within the box, the complete set of linearized equations of motion are given by

$$\ddot{x} = F_x + 3\Omega_k^2 x + 2\Omega_k^2 \dot{y} + \ddot{x}_m,$$

$$\ddot{y} = F_y - 2\Omega_k^2 \dot{x} + \ddot{y}_m, \qquad (4)$$

$$\ddot{z} = F_z - \Omega_k^2 z + \ddot{z}_m,$$

where $\mathbf{F} = -\nabla\phi$ is the sum of the gravitational and electrostatic forces per unit mass due to all other particles and the accelerations due to the magnetic field in each dimension are given by (Matthews *et al* submitted)

$$\ddot{x}_m = \frac{q}{m}\left[(\Omega_k - \Omega_p)R_x B_z + \dot{y}B_z - \dot{z}B_y\right]$$

$$\ddot{y}_m = \frac{q}{m}\left[\dot{z}B_x - \dot{x}B_z\right] \qquad (5)$$

$$\ddot{z}_m = \frac{q}{m}\left[-(\Omega_k - \Omega_p)R_x B_x + \dot{x}B_y - \dot{y}B_x\right].$$

2.2 Shepherding Satellites and Eccentric Orbits

The orbits of the F Ring, Prometheus, and Pandora all have small eccentricities [5,19]. While the orbits of the shepherding moons can be modeled accurately by box_tree as external gravitational potentials, the box containing the ring particles is at a fixed radial distance and thus models a circular orbit. To include the effect of the F Ring's eccentricity, the procedure of Showalter and Burns [7] is followed and the combined eccentricity of the rings and satellite is mapped back to the satellite alone. An "eccentric displacement vector" **E** is defined with polar coordinates ($ae$, $\omega$) where $a$ is the maximum radial displacement out-of-round for a given body, $e$ is the eccentricity, and $\omega$ is the longitude of the pericenter. Any ring/satellite configuration can thus be described by $a_0$, $a_1$, $\mathbf{E_0}$, and $\mathbf{E_1}$, where the subscripts zero and one refer to the ring and satellite, respectively. The effect of the satellite on the ring depends on $\mathbf{E_1} - \mathbf{E_0}$. By choosing to set $\mathbf{E_0}' = 0$ and $\mathbf{E_1}' = \mathbf{E_1} - \mathbf{E_0}$, all of the eccentricity can be assigned to the satellite. The effect of the "primed" satellite on a circular ring is essentially the same as that of the unprimed scenario, in which both the satellite and the ring have eccentric orbits. The reference longitude for the satellite's eccentricity is set at run time, so that the position of the satellite can be varied from closest approach to furthest approach as it passes the center of the box.

2.3 Boundary Conditions

The box_tree code makes use of periodic boundary conditions. In the general case, ghost boxes surround the central box and particles that are near a boundary interact with ghost particles in the neighboring box. As a particle leaves the central box, it is replaced by its ghost particle entering from the opposite side, keeping the total number of particles in the simulation constant. These boundary conditions were modified slightly to take advantage of special properties of the F Ring. Since the F Ring is a very narrow ring, the box size is chosen to be much greater than the radial width and vertical thickness of the ring. Thus particles are only able to leave the box in the azimuthal direction, requiring only two ghost boxes. Because the box is a non-inertial, rotating frame, particles that are displaced inward toward Saturn (-x direction) have a positive azimuthal drift (in the +y direction) while particles displaced outward (+x direction) have a negative azimuthal drift.

The shepherding satellites have different mean motions about the planet than does the box. Particles that are displaced inward (traveling faster than the mean motion of the box) will spend more time in the vicinity of Prometheus, the inner shepherding satellite. Likewise slower particles will spend more time in the vicinity of Pandora, the outer satellite. This effect was modeled by tracking the number of azimuthal boundary crossings, $n$, for each particle. The number of boundary crossings times the length of the box, $nL$, is added to the azimuthal position of the grain, and this adjusted position is used in calculating the grain-satellite separation and the gravitational force due to the satellite. The grain's unadjusted position within the box is used in calculating the effect of the magnetic field, which is azimuthally symmetric (Connerney 1993).

3. Results

For each run, the orbits of 5000 grains were tracked within a 2800 km box, which corresponds to 0.02 radians of arc at the distance of the F Ring. The grain radii followed a power law size distribution with q = 3.5 with minimum grain radius $a_{min}$ = 0.5 μm and maximum grain radius $a_{max}$ = 10.0 μm [5]. Grains were initially established within the 50 km F Ring envelope without perturbations from shepherding moons. Charged grains follow epicyclic orbits about a guiding center while uncharged grains have local Keplerian orbits. The plasma parameters in the F Ring are not well constrained, with estimates of plasma temperature ranging from 10 eV < $kT_e$ < 100 eV [6] and plasma density ranging from 10 cm$^{-3}$ < $n_o$ < 100 cm$^{-3}$ [21]. For this study, the plasma temperature and density were assumed to be constant across the width of the box. Using the values $kT_e$ = 20 eV, $n_o$ = 100 cm$^{-3}$, and dust density $N_d$ = 30 cm$^{-3}$ [21], calculated grain charges yield q/m = -0.03 C/kg for $a_{min}$ and q/m = -3.8×10$^{-6}$ C/kg for $a_{max}$ [23]. It should be noted that grain charges vary only slightly for $kT_e$ < 25 eV and as a result the relative dust to plasma density ratio has a relatively small effect on the qualitative results of this study [9].

Three types of ring/satellite interactions were modeled: Prometheus alone passing the ring section, Pandora alone passing the ring section, and Prometheus and Pandora passing the ring section simultaneously. The simulation was run using three consecutive boxes for each of the three cases with the data for the three concatenated to display a ring section that subtends 0.06 radians. The adjusted azimuthal positions were then plotted for grains that traveled faster than the mean motion of the box.

The initial ring configurations before and after passage of the moon(s) are shown in Figure 1 for both charged and uncharged grains. In all three cases with uncharged grains, the passing moon(s) excited an rippling orbit as shown in previous models (e.g. [4]) with the grains, across the entire size range, remaining in phase. However, in the case of the charged grains, a distinct phase difference evolves across different size populations with the larger charged grains (a ≥ 3.0 μm) having very similar orbits to the larger uncharged grains. Smaller grains (a < 1.0 μm), on epicyclic orbits displaced inward from the F Ring core, have increased orbital velocities and therefore are more strongly affected by Prometheus, the inner moon, and less strongly affected by Pandora, the outer moon. As a result, the orbital behavior of the charged and uncharged submicron grains differs considerably.

4. Conclusions

It has been shown that the presence of even very weakly charged grains within Saturn's F Ring can have a significant effect on the subsequent orbital behavior of the ring particles. Even when the grains are not charged strongly enough to display gravitoelectrodynamic resonances with the shepherding moons, the differing charge-to-mass ratios over the various size grains will lead to phase differences in the rippling orbits imposed on the ring particles by the passing moons. This effect may in turn contribute to the observed braiding in the F Ring.

This model investigated only the effect of Saturn's gravitational and magnetic fields along with the gravitational fields of the shepherding satellites. It is presently computationally infeasible to model both the large-scale structure of the F Ring as well as a realistic number density; such a model would require on the order of $10^{16}$ particles. Accordingly, the dust number density modeled in this system does not correspond to the actual dust number density for the F Ring and as such particle interactions and collective effects of the charged dust grains and the plasma were ignored. However, since the primary goal of this work was to examine the perturbation to individual grain orbits due to their interaction with Saturn's magnetic field, a realistic number density was used in calculating the grain charges.

While the dust density used in this model is too low to model the collective effects of the charged dust grains and the plasma, it is very interesting to note that distinct dusty plasma clouds repel each other due to interaction of the plasma sheaths at the boundaries (Morfill, personal communication 2002). If the phase differences created by the shepherding moons truly separate the dust into different populations according to size, it is possible that these populations would resist recombination, helping to preserve distinct ringlets. The arrival of Cassini in 2004 will provide an opportunity to investigate the structure of the F Ring and its plasma environment in detail. It would be instructive to learn if separate strands have differing size populations, or if there is a size gradient across the width of the F Ring envelope. Such size sorting would confirm the presence of charged grains within the F Ring and underscore the importance of including even very weak charge effects in modeling ring structures.

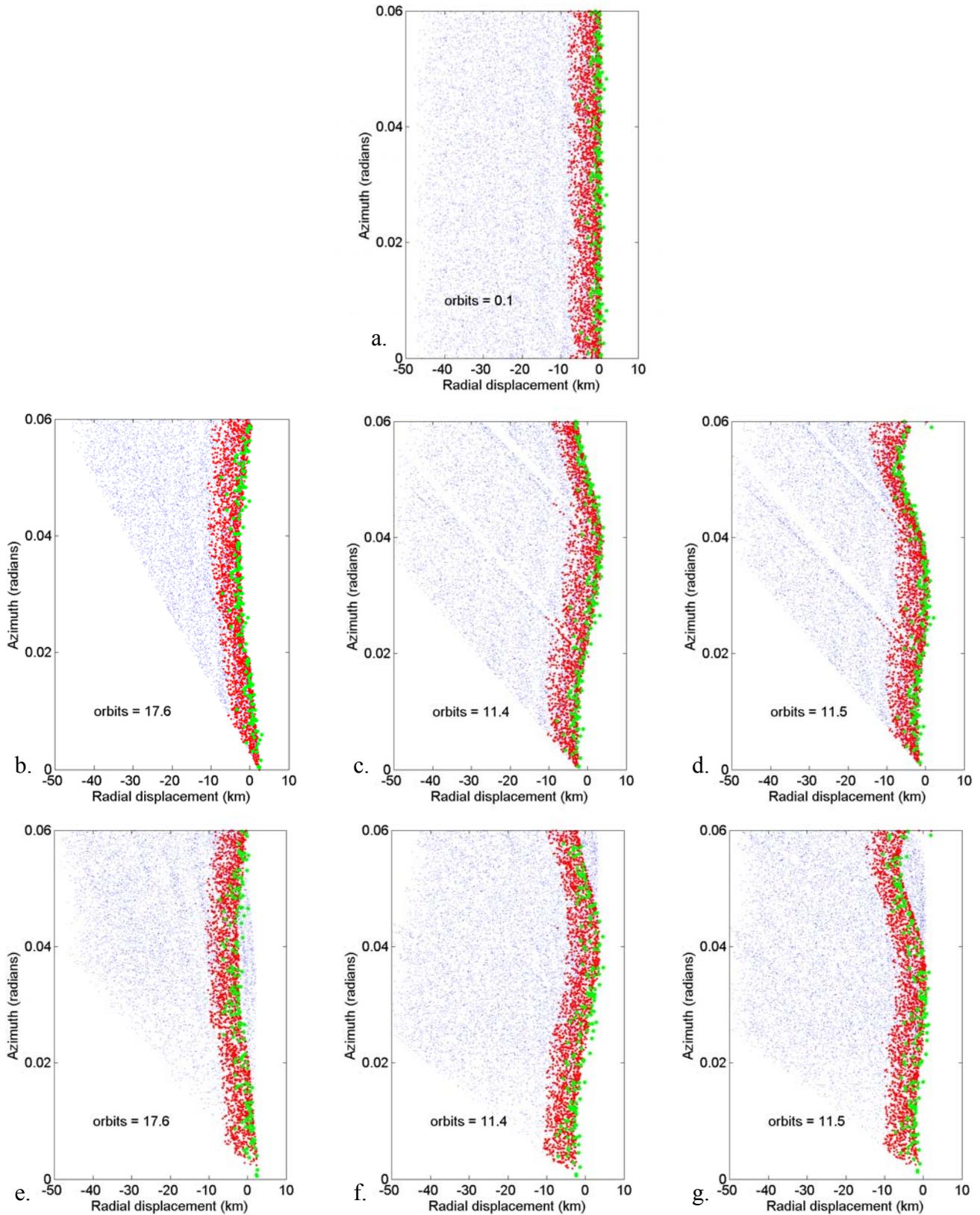

Figure 1. Locations of grains within the F Ring after passages of the shepherding satellites. Grains with r < 1.0 μm are shown in blue, grains with 1.0 μm ≤ a < 3.0 μm are shown in red, and grains with r ≥ 3.0μm are shown in green. The initial configuration of the ring particles is shown in (a). The smaller charged grains have larger epicyclic orbits due to their higher charge-to-mass ratios. The number of orbits refers to the number of revolutions of the center of the box about Saturn. (b)-(d) Uncharged grains after passage of Prometheus, Pandora, and both moons simultaneously, respectively. (e)-(g) Charged grains after passage of Prometheus, Pandora, and both moons simultaneously, respectively.